\documentclass[twocolumn]{aa}
\usepackage{graphicx}
\usepackage{txfonts}
\def\hs{HS\,0922+1333}
\def\asec{\ifmmode ^{\prime\prime}\else$^{\prime\prime}$\fi}
\def\etal{{et\,al. }}
\def\msun{M$_{\odot}$}

\def\mdot{$\dot M$}

\def\it{\sl}
\def\degs{\ifmmode ^{\circ}\else$^{\circ}$\fi}
\def\amin{\ifmmode ^{\prime}\else$^{\prime}$\fi}
\def\asec{\ifmmode ^{\prime\prime}\else$^{\prime\prime}$\fi}
\def\fd{\hbox{$.\!\!^{\rm d}$}}            
\def\farcs{\hbox{$.\!\!^{\prime\prime}$}}  

\def\degs{\ifmmode ^{\circ}\else$^{\circ}$\fi}
\def\amin{\ifmmode ^{\prime}\else$^{\prime}$\fi}

\def\eqalign#1{\null\,\vcenter{\openup1\jot \m@th
   \ialign{\strut\hfil$\displaystyle{##}$&$\displaystyle{{}##}$\hfil
   \crcr#1\crcr}}\,}
\begin{document}
   \title{On the orbital period of the magnetic Cataclysmic Variable HS\,0922+1333}


   \author{G. H. Tovmassian
          \thanks{Visiting
research fellow at Center for Astrophysics and Space Sciences,
University of California, San Diego, 9500 Gilman Drive,
La Jolla, CA 92093-0424, USA}
          \and
          S.V. Zharikov
          }

   \offprints{G. Tovmassian}

   \institute{Observatorio Astr\'onomico Nacional SPM, Instituto de Astronom\'{\i}a, Universidad Nacional 
Aut\'onoma de M\'exico, Ensenada,
BC, M\'exico\thanks{PO Box 439027, San Diego, CA, 92143-9024, USA};\\
\email{gag,zhar@astrosen.unam.mx}
                     }

   \date{}

  \abstract{The object \hs\ was visited briefly in 2002 in a mini survey of
low accretion  rate polars (LARPs) in order to test if they undergo high luminosity states
similar to ordinary polars. On the basis of that short observation the suspicion arose that
 the object might be an asynchronous polar (Tovmassian \etal
 2004). The disparity between the presumed orbital and spin
 period appeared to be quite unusual. }
 {We performed follow-up  observations of the object to resolve the problem. }
 {New simultaneous spectroscopic and photometric observations spanning several years
 allowed measurements of radial velocities of emission and absorption lines from the secondary star 
 and brightness variations due to synchrotron emission from the primary.}
 {New observations
 show that the object is actually synchronous and its orbital and
 spin period are equal to 4.04 hours.}{ We identify the source of confusion of previous observations to 
be a high velocity component of emission line arousing from the stream of matter leaving L$_1$ point.}{ 
   
 }
   
  \authorrunning{Tovmassian \& Zharikov}
  \titlerunning{Orbital period of the HS\,0922+1333}

   \maketitle
%

\section{Introduction}

Magnetic cataclysmic variables (CV) are accreting binary systems in
which material transfers from a dwarf secondary star onto a magnetic
($\sim$5 $<$ B $<$  $\sim$250 MG) white dwarf (WD) through Roche lobe
overflow. Polars or AM\,Her systems with magnetic fields larger than
$\sim10$ MG stand out among magnetic CVs because the spin period of the primary WD
is synchronized with the orbital  period of the system. Unlike
non-magnetic or low-magnetic accreting binaries, they have neither 
a disk nor the capacity to accumulate the transferred matter, so the 
bulk of flux of these systems comes from the accretion
flow,  particularly around magnetic  poles. Therefore, their luminosity is
sensitive to the mass transfer rate \mdot.  Polars are known to have
highs and lows in their luminosity state, which is directly dependent
on \mdot.

In recent years a number of polars were identified with extremely
 low accretion rates. They are commonly called LARPs, a name coined by
 Schwope \etal (\cite{schwope02}).  Their mass accretion rate is
 estimated to be about a few $10^{-13}$ \msun/yr, two orders of
 magnitude below the average for CVs and they are distinguished for
 their prominent cyclotron emission lines on top of otherwise
 featureless blue continua. The first two LARPs, including the subject of this
 study, were discovered in the course of the Hamburg QSO survey, thanks to
 a broad variable feature in the spectra subsequently  identified with
 cyclotron lines (Reimers \etal \cite{rehaho99}; Reimers \& Hagen \cite{reha00}, hereafter RH). 
Later, another newly  identified magnetic CV from the list of ROSAT sources (RX\,J1554.2+2721) was
 spotted in the low state with a spectrum identical to LARPs (Tovmassian \etal
 \cite{to01,to04}).  Intrigued by that discovery, we conducted a blitz
 campaign to check if canonical LARPs, namely HS\,1023+3900 and
 HS\,0922+1333, might be caught in a high state as well. 
 
 Since both objects had only recently been discovered and had very limited observational
 coverage, we obtained one full binary orbital period of spectral
 observations. Our instrumental setup  provided higher spectral resolution than
 the original discovery observation by RH. We observed emission from the
 H$\alpha$ line apparently arising from the irradiated surface of the
 secondary star facing the hot accreting spot on the WD 
 and Na\,{\sc i} infrared doublet from the cooler parts of the
 secondary star (Tovmassian \etal \cite{to04}). The derived radial velocity (RV) curve
 from that observation  did not fold well with the period
 estimated in the discovery paper. However, the limited time coverage
 undermined our ability to measure the period properly. 
 We could only state that the period might be exceeding what was reported by RH by   
at least 1.14 times, corresponding to P$_{\mathrm {spin}}$/P$_{\mathrm {orb}}$=0.88. 
It should be noted that RH  determined
 their period from the cyclotron hump cycles and thus, they measured 
 the WD spin period rather than the binary orbital period.  It would be 
quite usual to find some degree of
 de-synchronization between the spin period of the WD and orbital
 period. Nevertheless, the difference in periods was too large for an
 asynchronous polar and too small for an intermediate polar. The latter mostly  
follow the empirical ratio  P$_{\mathrm {spin}}$/P$_{\mathrm {orb}} \sim0.1$. 
In rare cases, P$_{\mathrm {spin}}$/P$_{\mathrm {orb}} \sim 0.25$
 (see e.g. Norton \etal \cite{norton}).
 There are also theoretical restrictions on a kind of ratio that was
 indicated by our observation as evident from the Norton \etal (\cite{norton})
 paper.  Therefore, we conducted a new series of observations in order
 to establish the orbital period of the compact binary and to 
 classify it properly.  This brief paper analyses a combined set of
 observations and discusses the reasons that led us to an erroneous conclusion
 in \cite{to04}. 

In  Sect.\ref{Obs} we  describe our  observations and  the data reduction.
The data  analysis and the  results are presented  in Sect.\ref{DatAn}, 
and conclusions are drawn in Sect.\ref{Summ}.

\begin{table*}[]
\begin{center}
\caption{Log of observations of \hs}
\begin{tabular}{llllclc}
\hline\hline
 Date     & HJD+ & Telescope& Instrument/Grating  &Range/Band & Exp.Time & Duration \\ 
Spectroscopy &  24530000       &   &        &     & Num. of Integrations&         \\ 
2002 02 04 & 2309 &2.1m &B\&Ch$^1$ 600l/mm  & 6200-8340\AA    &  900s$\times$19   & 4.5h  \\
2003-03-25 & 2723 &2.1m &B\&Ch\ \ \ 1200l/mm  & 6100-7200\AA    &  900s$\times$13   & 3.3h  \\
2003-03-26 & 2724 &2.1m &B\&Ch\ \ \ 1200l/mm  & 6100-7200\AA    &  900s$\times$6   & 1.3h  \\
2003-03-27 & 2725 &2.1m &B\&Ch\ \ \ 1200l/mm  & 6100-7200\AA    &  900s$\times$15   & 2.6h  \\
2005-10-27 & 3670 &2.1m &B\&Ch 400l/mm   & 6100-9200\AA    &  900s$\times$15  & 2.2h  \\
2005-10-29 & 3672 &2.1m &B\&Ch 400l/mm   & 6100-9200\AA    &  900s$\times$11  & 1.7h  \\
2005-10-30 & 3673 &2.1m &B\&Ch 400l/mm   & 6100-9200\AA    &  900s$\times$8  & 1.2h  \\
2006-01-18 & 3673 &2.1m &B\&Ch 400l/mm   & 5800-8900\AA    & 1800s$\times$8  & 3.7h  \\
Photometry &  &   &    & &   \\
2003-03-26 & 2724 &1.5m &  RUCA$^2$  &   R    &  120s$\times$107      & 3.8h  \\
2003-03-27 & 2725 &1.5m &  RUCA      &   R    &  120s$\times$99     & 3.4h  \\          
2003-03-28 & 2726 &1.5m &  RUCA       &  R    &  120s$\times$99     & 3.4h  \\          \hline               
\end{tabular}
\label{tab1}
\end{center}
\begin{tabular}{l}
$^1$ B\&Ch - Boller \& Chivens spectrograph (http://haro.astrospp.unam.mx/Instruments/bchivens/bchivens.htm) \\
$^2$ RUCA - CCD photometer (http://haro.astrospp.unam.mx/Instruments/laruca/laruca\_intro.htm)
\end{tabular}
\end{table*}

\section{Observations and reduction}
\label{Obs}

Sets of observations were collected over a four-year period  and analyzed.
All observations of \hs\ reported here were
obtained at the Observatorio Astr\'onomico Nacional San
Pedro Martir, M\'exico. The B\&Ch spectrograph installed at the 2.1 meter telescope was
used for the extensive spectroscopy, while a 1.5\,m telescope was used to 
obtain simultaneous photometry during the 2003 March run. 
In the first observations, upon which Tovmassian \etal (\cite{to04}) depended, 
we used a 600 l/mm grating centered
in the optical IR range (6200 -- 8340 \AA) to achieve a spectral resolution of 
4.2\,\AA\ FWHM  in a sequence of 900\,sec exposures covering one orbital period.
The controversy over the periods led us to re-observe the object during three nights 
in March 2003. 
This time we  utilized the highest available grating of
1200\,l/mm. The spectral resolution reached 2.2\,\AA\,  FWHM covering the 
6100 -- 7200\,\AA\ range. Later we collected more observations with lower
resolution to refine the orbital period and properly classify the secondary star.

In all observations an SITe $1024\times1024$ 24~$\mu$m pixel CCD was used to acquire the data.
The slit width was usually set to 2\farcs0 and oriented in the E--W direction. He-Ar arc lamp 
exposures were taken at the beginning and end of each  run for wavelength
calibration. 

In 2003 March observations we conducted simultaneously with differential photometry.
Exposure times were  40--60 sec 
with an overall time resolution of about  80--100 sec using the Johnson-Cousins R$_c$ filter. 

The reduction of data was done in a fairly standard manner. The bulk of
reduction was performed using IRAF\footnote{http://iraf.noao.edu} procedures, except for
removing of cosmic rays by a corresponding program in  MIDAS\footnote{ESO-MIDAS is the acronym for the 
European Southern Observatory Munich Image Data Analysis System which is 
developed and maintained by the European Southern Observatory}, as this is an
easier and more  reliable tool. The biases were taken at the beginning
and end of the night and were subtracted after being combined  using the
CCD overscan area for control of possible temperature-related
variations during the night. We did not do flat field correction for
spectral observations and used blank sky images taken at twilight for
direct images. The flux calibration was done by observing a
spectrophotometric standard star.  Feige\,34 was observed during a 2002
run and G191-B2B during the rest of the observations. 

The wavelength calibration is routinely done by observing a
He--Ar arc lamp at the beginning and end of a sequence on the object or
every 2 hours if the sequence is too long. Then the wavelength solutions
calculated for each arc-lamp exposure and an average of
preceding and succeeding images are applied to the object observed
in between. The wavelength solutions are usually good to a few 1/10
 of an Angstrom, while deviations due to the telescope
position and flexations of the spectrograph can exceed that by an
order of magnitude. Usually, that does not pose a problem since we work
with moderate resolutions and the amplitude of radial velocity variation
is on the order of hundreds of km/sec. The sensible way of checking and
correcting wavelength calibration is to measure the night sky
lines. We measured several lines by selecting unblended ones located
close to H$\alpha$ and Na\,{\sc I}.
The measurements of sky lines show a clear trend and indicate the scale of 
errors that one can incur 
depending on the telescope inclination.  Although
the trend is unusually steep, reaching 30\,km/sec over 4 hours of observation,
the scatter of points around a linear fit is relatively small, which
defines the error of the measurements (rms) and is  $\leq$8 km/sec. 
Nevertheless, the error bars in the corresponding plots 
reflect the entire range of deviation just to
demonstrate the scale of corrections applied to the data. The
deviations of the linear fit to the measured night sky lines 
(with an average of 2 night sky lines around each
measured line) from the rest value were used to correct the wavelength
calibration by the corresponding amount.

\begin{figure}[t]
\includegraphics[width=90mm, clip=]{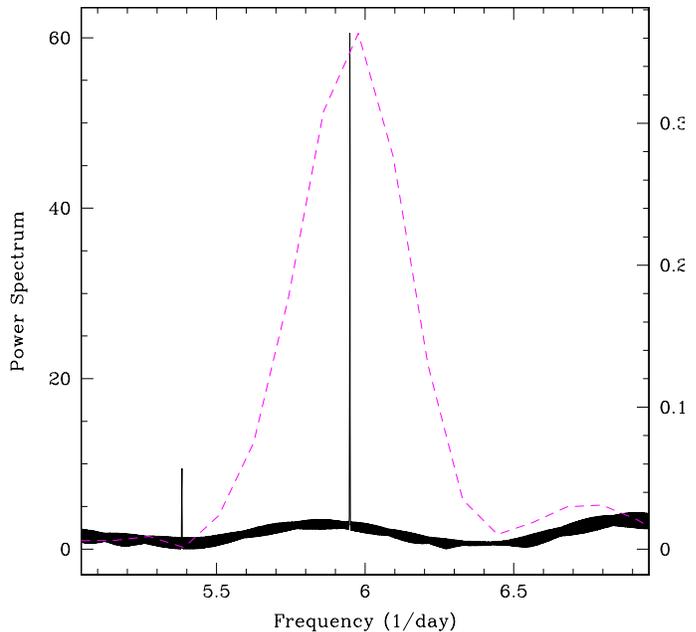}
\caption{The CLEANed power spectrum of the RV variation is presented by a solid line.
The dashed line is the power spectrum of photometric data. Vertical axes on the right side correspond to the
photometric power scale.}
\label{fig1}
\end{figure}

\begin{figure}[t]
\includegraphics[width=81mm, clip=]{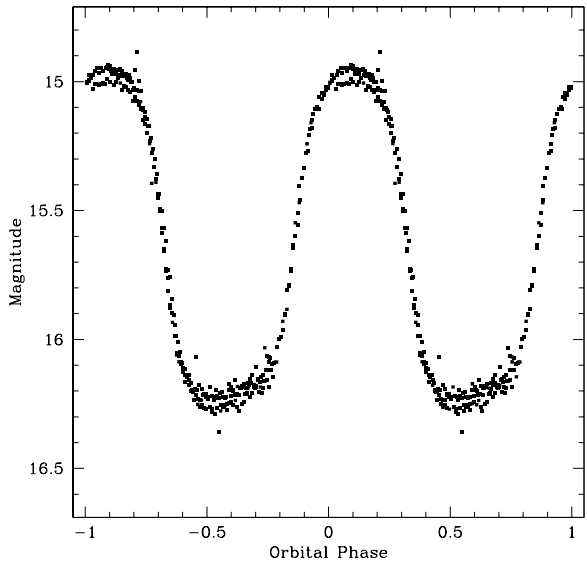}
\caption{The light curve of \hs\ obtained in filter R$_c$ and folded with the orbital period. The phasing is according to spectroscopic data. }
\label{fig2} 
\end{figure}

\section {Orbital period and system parameters}
\label{DatAn}

We measured the H$\alpha$ line in the 2002 spectra with single Gaussian fits. The resulting RV
curve was reasonably smooth and sinusoidal, but the ends of the curve would not 
overlap when folded with the period
reported by RH (see Fig. 5 in \cite{to04}). We speculated
that the actual orbital period is longer than the one derived from the
photometry. However, the measurements of new spectra obtained in 2003
do not show such a large discrepancy, and the period analysis of the
combined dataset easily reveals that the true period is indeed 4.0395
hours and coincides with the photometric period derived from the
synchrotron lines  variability within errors of measurement.  The combination of data taken years
apart and several nights in a row each year allowed  us to determine the period
very precisely. We applied the CLEAN procedure (Roberts \etal \cite{clean}) to sort out
the alias periods resulting from the uneven data sampling and daily
gaps and obtained a strong peak in the power spectrum at the
$5.94131\pm0.00065$ cycles/day, corresponding to a $4.0395\pm0.0001$
hour period (see Fig. \ref{fig1}).  Simultaneous with spectroscopy, we obtained photometry in
R$_c$ band that partially includes the strongest cyclotron line. It
is a dominant contributor to the light curve (Fig.\ref{fig2}), so we can use
it to determine the spin period of the WD.  The power spectrum
calculated from photometry gives exactly the same result, but the peak
is broader, because the data lacks a longer time base. 
In  Fig.\ref{fig1} the power spectra of spectral and
photometric data are presented together.

\begin{figure*}[t]
\vspace*{0.2cm}
\setlength{\unitlength}{1mm}
\resizebox{8cm}{!}{
\begin{picture}(100,100)(0,0)
\put (10,0){\includegraphics[width=100mm, clip=]{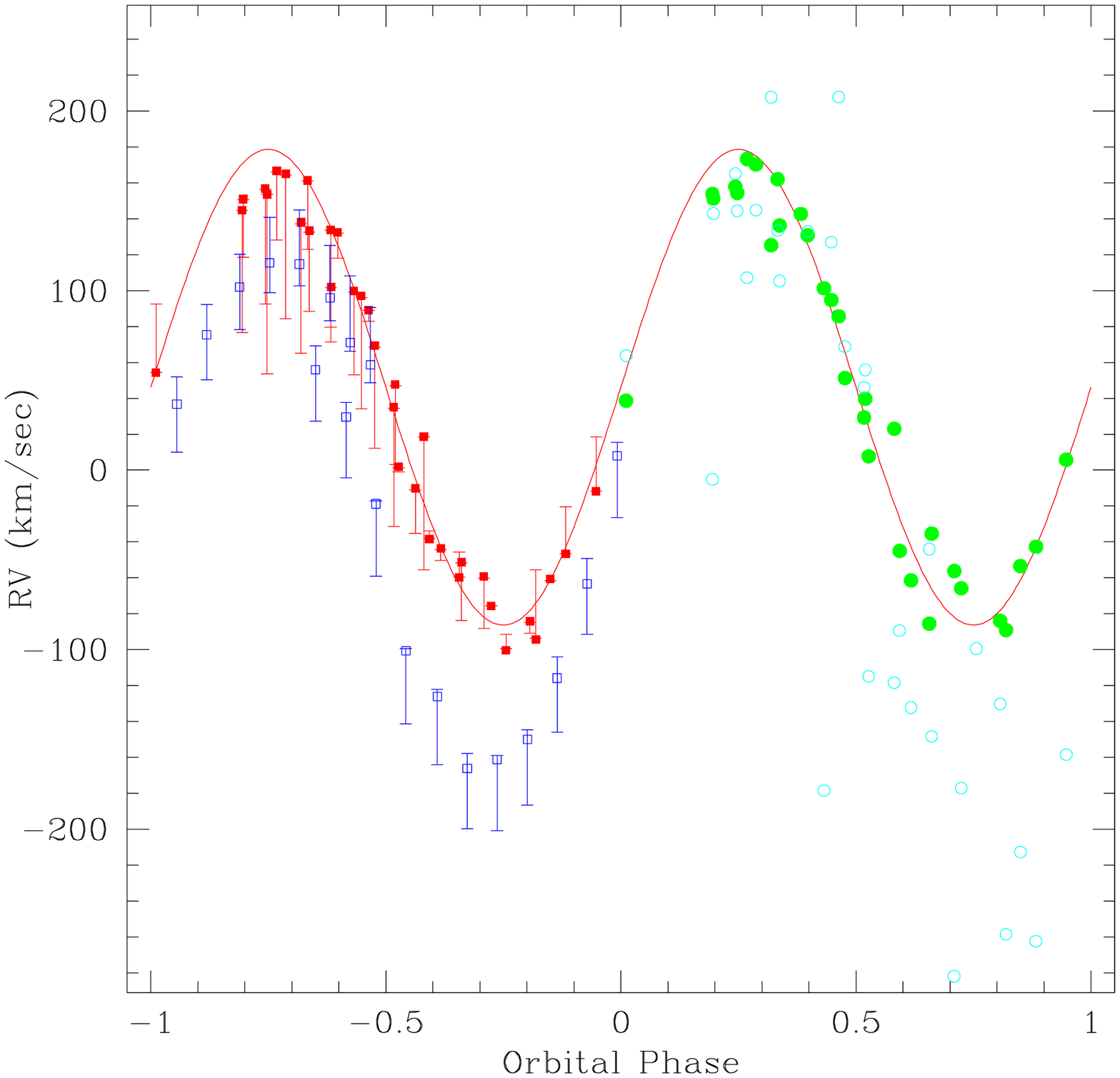}}
\put (110,0){\includegraphics[width=100mm, clip=]{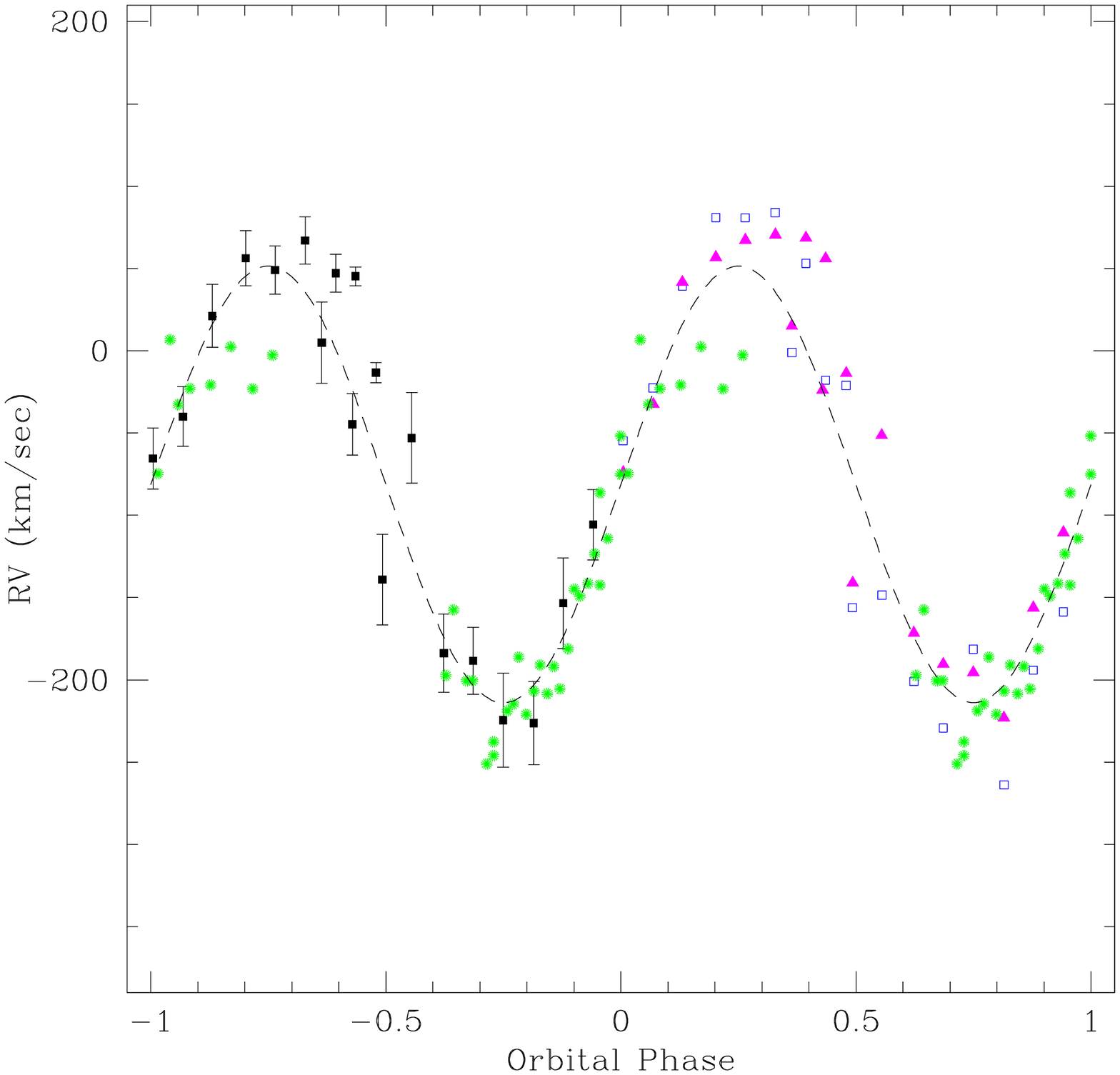}}
\end{picture}}
\caption{The radial velocity curve of H$\alpha$ line (left panel) and
of Na\,{\sc i} doublet (right). The open squares (blue) in the left panel represent 2002
data obtained with lower spectral resolution. The filled squares (red)
are measurements of 2003 observations with single Gaussian fitting. The
error bars on the left side of the plots reflect the amplitude of
wavelength  corrections. The points are placed at the correct positions after trend removal. The
right side of the plot presents measurements of the  2003 data but
with double Gaussian de-blending of the line. The filled (green)
circles are from the stronger line component originating at the
irradiated secondary, the open circles correspond to a much weaker component
coming  from the stream. 
In the right panel, measurements of  the \ion{Na}{I} lines are presented
from 2002 observations.  The filled square symbols denote RV of $\lambda\,8197$ \AA\ measured with Gaussian 
deblending, after velocity correction with sky lines. The open squares and triangles are 
measurements of the same doublet with single gaussians (squares $\lambda$\,8185\AA\  and triangles  $\lambda$\,8197\,\AA).  
The diamonds are measurements of the $\lambda$\,8185\,\AA\ line from 2006 observations.
The  curve is a result of  $sin$ fit to the combined data.
Note that scales of y-axes of panels are different.}
\label{fig3}
\end{figure*}

It is clear that this system is a synchronous magnetic cataclysmic
variable. The spin period of its white dwarf primary is locked with the 
orbital
and is not shorter, as suspected earlier. We explored 
the cause of confusion. First of all, we corrected all measured radial velocities
using the night sky lines to remove the trends. This decreased the gap a little 
between points in the 2002 data in
phases 0.0 through 0.2 where they were not overlapping. 
But even taking errors related to the wavelength calibration  into account,
they still do not fold properly (see the open (blue) square symbols in
Fig.\ref{fig3}). What is more interesting, however, is that the
amplitude of the radial velocity variation has a much higher value in the 2002 
data than in 2003 data, as measured with single Gaussians. 
The careful examination of the 2003 data, with
twice the spectral resolution than in 2002, reveals that at the bottom of the
H$\alpha$ emission line there is a weak and broad bump present in most
 phases. We  de-blended the H$\alpha$ line from 2003
observations using two Gaussian components in the IRAF {\it splot}
procedure.  The result is shown on the right side of
the left panel of Fig.\ref{fig3} only (positive phases). The strong, narrow component  
basically coincides with the single Gaussian
measurements. But the weak broad component appears to show a much larger
amplitude and reveals itself mainly between phases 0.3 through
0.9. This component is clearly identified as the heated matter
leaving the Lagrangian L$_{1}$ point, the nozzle where the accretion
stream forms. Outflowing matter has intrinsic velocity, so at phase 0.75 when 
the secondary star reaches maximum velocity toward the
observer,  it tilts the weight of the emission line toward
larger velocity. Its phasing 
appears to be similar to a high-velocity component (HVC) detected routinely
in polars (Schwope \etal \cite{schwope}, Tovmassian \etal \cite{tov}) that originates in the ballistic part of the stream. 
In the lower-resolution spectra  this component could not be separated, therefore the
radial velocity curve became stretched and deformed. That and the
short time coverage limited to just a little over one orbital period
led to the misinterpretation of the 2002 spectral data. It is
very interesting that we were able to distinguish the accretion flow
onset. So far these objects have been known to show only the synchrotron
humps as an evidence of accretion processes taking place in them (Schwope \etal \cite{schwope02}).

The RV curve derived from sodium lines
(see Fig.\ref{fig3})  gives the measure of the
rotation of the center of mass of the secondary in the orbital plane,
while the narrow component of the H$\alpha$ line originates from the front side of the
elliptically distorted secondary. 
The ephemerides of \hs\  from the  RV measurements can be described as

$$ \rm T_{\rm 0} = {\rm HJD}\, 2452308.336 + 0\fd168313[200] \times {\rm E},  $$ where $ \rm T_{\rm 0}$ 
corresponds to 
the $-/+$ crossing of the RV curve as follows from the fitting of sinusoid to the RV measurements of  H$\alpha$ 
and sodium lines separately according to the following equation:
 
$$ V(t) = \gamma + K\times sin(2\pi (t-t_0)/P_{orb} ) $$

Corresponding numbers derived from the fitting are presented in the Table \ref{rvtab}. Unfortunately,  due to the large errors there is no marked difference between the semi-amplitude of radial velocities between 
H$\alpha$ and  \ion{Na}{I} lines. Otherwise, knowing the spectral type of the secondary, we could deduce the basic parameters of the binary since that difference reflects the size of the Roche lobe of the secondary. 


\begin{table}
\begin{center}
\caption{Radial velocity parameters of \hs.}
\label{rvtab}
\begin{tabular}{lccc}
\hline\hline
 Line &  $\gamma $    &   K     & Residuals   \cr
 &  km/sec  &  km/sec &   km/sec  \cr \hline
H$\alpha$ &  36.6$\pm$7 & 132$\pm$12 &  25.1 \cr
\ion{Na}{I} 8185\AA & -81$\pm$11 & 162$\pm$17 &  29.5  \cr
\ion{Na}{I} 8197\AA & -65$\pm$13 & 139$\pm$20 &  32.7 \cr \hline
\end{tabular}
\end{center}
\end{table}

\begin{figure}[h]
\includegraphics[width=90mm, clip=]{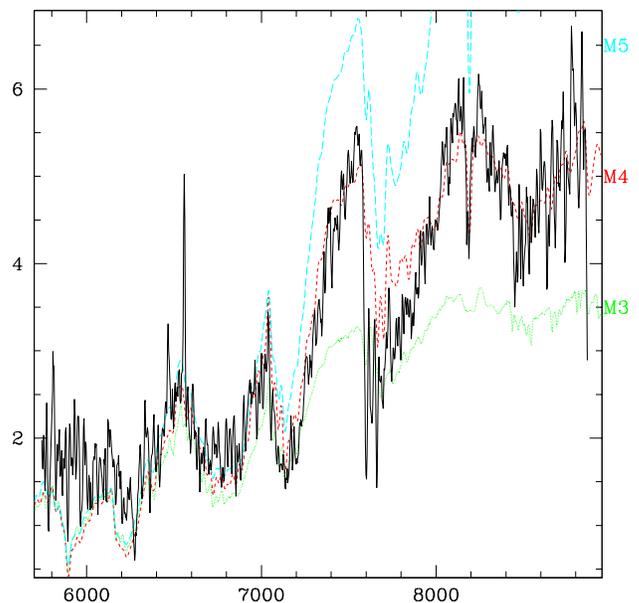}
\caption{ The spectrum of \hs is presented by the solid line. For comparison the standard spectra
of M3-M5 stars are plotted from the Pickles (\cite{uksp})}
\label{fig3c}
\end{figure}

The spectrum of the
secondary in the absence of an accretion disk is clearly seen, and in the
phases when the magnetic accreting spot that is radiating strong synchrotron
emission is self-eclipsed, one can see undisturbed secondary spectrum
in the near infrared range. In the Fig. \ref{fig3c} the flux calibrated 
spectrum of the object obtained at phase 0.5 is presented. Overplotted are standard 
spectra of M3 to M5 main sequence stars (Pickles \cite{uksp}) normalized to the object.
The WD's contribution has not been removed. However, at wavelengths above 6500\,\AA,
its contribution is apparently insignificant and a good accordance emerges between the object and
M4\,V standard star. 
This is also consistent with what is
expected from the P$_{\rm orb}$ - spectral type II relation
(Beuermann \cite{beu}), although the secondary is a M3.5 star according to RH. 
The masses of secondaries in systems with periods similar to \hs\ range 
from 0.35 to 0.42 \msun, in those cases where the mass could be estimated 
precisely. 
Such a secondary would follow the empirical mass-period and radius-period
relations from Smith \& Dhillon (\cite{smdhi}) 
\begin{eqnarray}
M_2/M_\odot & = & 0.126(11)\ P(h) - 0.11(4) \nonumber \\
R_2/R_\odot & = & 0.117(4)\ P(h)-0.041(18),
\end{eqnarray}

Observations with higher resolution in the near IR will permit investigators to precisely measure 
the difference between the RV of H$\alpha$ originating at the facing side of the secondary and sodium 
absorption lines reflecting the
motion of the center of mass. Subsequently, it should allow for  estimating the observed 
radius of the star to check the possibility that it fills the Roche lobe. For now, we can only 
assume that the mass transfer proceeds in a way similar to other polars, based on the 
detection of a high velocity component in the emission line. Its presence can also be 
illustrated by constructing Doppler tomograms.

\begin{figure}[t]
\includegraphics[width=90mm,bb=150 220 555 660, clip=]{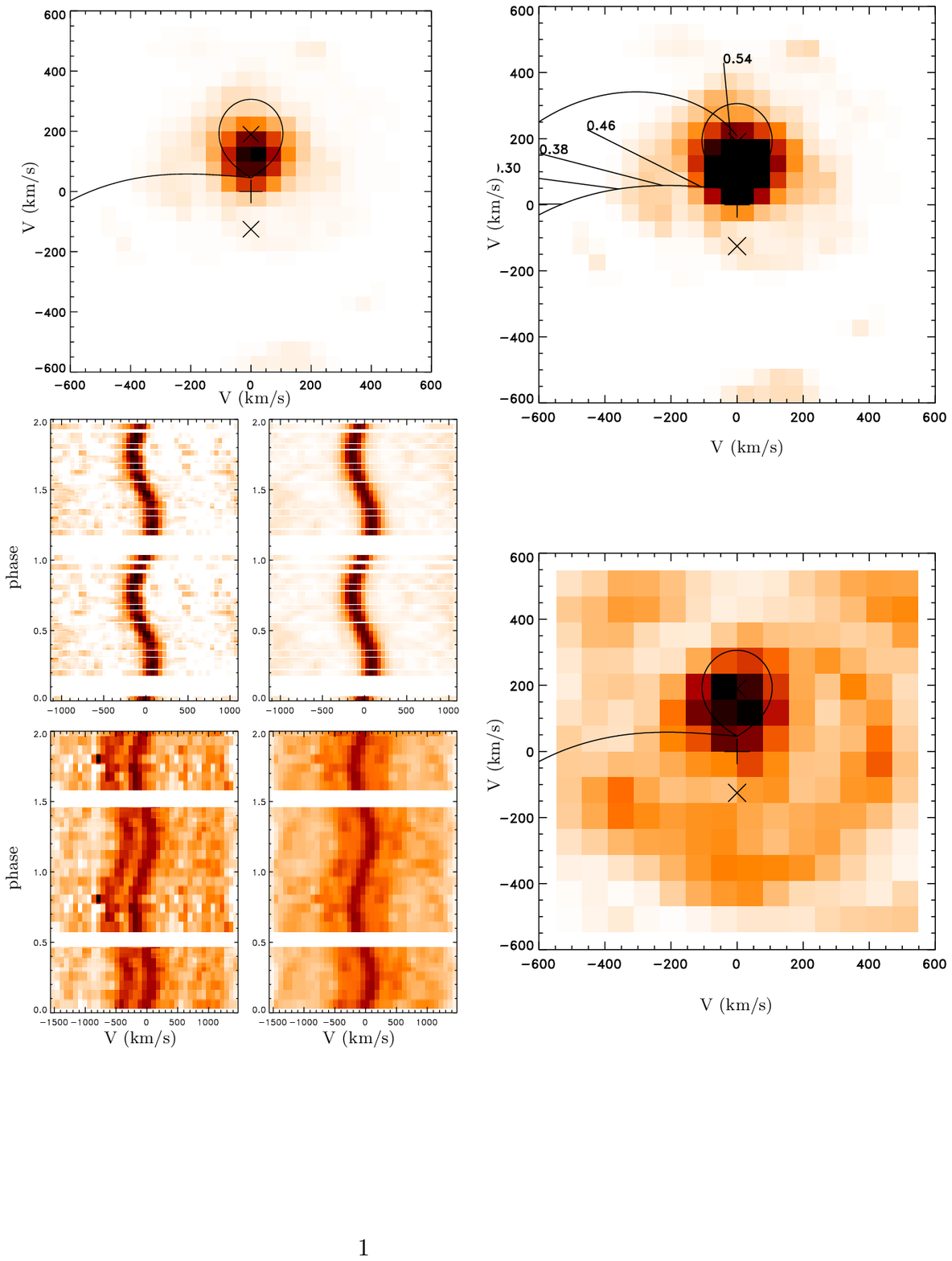}
\caption{The Doppler maps of \hs. On the top the tomograms of H$\alpha$ emission line are presented 
in two panels with different contrast levels to emphasize the concentration of the emitting region on the facing side of the secondary on the left and possibly some trace of mass transfer stream on the right. The curved lines in the top panels correspond to the stream trajectory, with numbers in the top right panel indicating stream azimuth.
The tomogram corresponding to the \ion{Na}{I} line is placed below in the right corner.  The circle-shaped emission around the center of mass is caused by the presence of the component of the doublet line.
In the bottom left corner, the observed and reconstructed trailed spectra of  H$\alpha$ line (above) and  \ion{Na}{I} (below) are presented.}

\label{tomography}
\end{figure}

Doppler tomography (Marsh \& Horne
\cite{Marsh88}; Marsh \cite{Marsh02}) is a powerful tool in cases like
this, where the origin of line profiles is bound to the orbital plane
and the system has relatively high inclination. We constructed
Doppler maps, or tomograms, using both the H$\alpha$ emission line and
the Na\,{\sc I} $\lambda 8197 \AA$ absorption line to prove the
accuracy of our estimate of the binary parameters. The
tomograms in Fig.\ref{tomography} show that the H$\alpha$ line is mostly
confined to the front side of the secondary, while the sodium
absorption fills the entire body of the secondary. However, the
difference is not very obvious. The reason for that appears to be the
 the lower spectral resolution and fewer spectra employed.

\section{Conclusions}
\label{Summ}

   \begin{enumerate}
      \item We have determined the 4.0395 hours spectroscopic period of the
      LARP \hs\ based on the radial velocity measurements of H$\alpha$ emission
line originating at the irradiated secondary star. The derived value coincides within
      measurement errors with the spin period of the system, thus
      proving that the object is a synchronized polar.
         \item The profiles of the H$\alpha$ emission line in higher-spectral 
         resolution observations turned out to be complex. They are
         formed  basically on the irradiated surface of the
         secondary star, but they also show a small contribution from the matter
         in close proximity to the L$_1$ point. The matter escaping the secondary 
         shows RVs with higher velocity and a different phase.
        \item The Doppler tomograms tend to confirm detection of a stream of transfer matter.
           \end{enumerate}

The parameters of the system that we have obtained are interesting in the context
of the model proposed by Webbink and Wickramasinghe
(\cite{ww2005}). According to it, the LARPs are relatively young and
are still approaching their first Roche lobe overflow. The accretion is
due to the capture of the wind material from the secondary by the strong
magnetic field of the primary. We think that we see evidence of a faint stream common to
the polars that transfer material through the L$_1$ point which is usually due to the Roche
lobe overflow. However, the wind will probably also cause a flow of matter through the same 
trajectory, so it is difficult to say if the observation runs against the model. The precise 
measurement of the secondary star size may help to clarify this.

\begin{acknowledgements}
  This study was supported partially by  grant 25454 from CONACyT. GT acknowledges the
UC-MEXUS fellowship program enabling him to visit CASS UCSD. The authors are grateful
to the anonymous referee for careful reading of the manuscript and valuable comments. We
thank L.Valencic for help in language related issues. 
      \end{acknowledgements}

\end{document}